\documentclass[runningheads]{llncs}
\usepackage[T1]{fontenc}
\usepackage{graphicx}
\usepackage{subcaption}
\usepackage{array}
\usepackage[table]{xcolor}
\usepackage[flushleft]{threeparttable}
\usepackage{amssymb}
\usepackage{amsmath}
\usepackage{hyperref}
\usepackage{color}
\usepackage{orcidlink}
\usepackage{bbding}

\begin{document}

\title{Cyber-Electromagnetic Anomaly Detection Through Time-Series Analysis}

\author{María Teresa Guillén Navarro\orcidlink{0009-0009-8531-3981}\inst{1} \and Juan Luis Serradilla Tormos\orcidlink{0009-0009-3382-5272}\inst{1} \and Sergio López Bernal\orcidlink{0000-0003-1869-1965}\inst{1} \and Daniel Orlando Díaz López\orcidlink{0000-0001-7244-2631}\inst{1}\Envelope \and Gregorio Martínez Pérez\orcidlink{0000-0001-5532-6604}\inst{1}}

\authorrunning{M. T. Guillén Navarro et al.}

\institute{Department of Information and Communications Engineering, University of Murcia, Campus de Espinardo, 30100, Murcia, Spain\\
\email{\{mt.guillennavarro, juanluis.serradillat, slopez, danielorlando.diaz, gregorio\}@um.es}}

\maketitle            
\begin{abstract}

Military operations benefit from the coordination between kinetic and non-kinetic domains. In particular, the coordination of cyber operations and electromagnetic warfare has become increasingly relevant for gaining operational advantage. This coordination is also relevant for Cyber Situational Awareness (CSA), where the Observe-Orient-Decide-Act (OODA) loop requires monitoring and interpreting evidence from heterogeneous sources. In this context, anomalies may appear not only in the physical behavior of signals, but also in the communication behavior observed at the traffic level. However, many existing anomaly detection proposals focus on only one of these perspectives, limiting their ability to characterize events that manifest simultaneously in the electromagnetic spectrum and cyberspace. To address this limitation, this work develops and evaluates two anomaly detection models that combine features from both domains. More specifically, the study uses the ZBDS2023 dataset, which contains traffic from nodes in a mesh network, including benign and attack behaviors. Thus, this dataset provides physical-level features, traffic-level features, and labeled attacks. Two detection approaches are evaluated: a supervised model based on Random Forest and an unsupervised model using LSTM-Autoencoder. The results show that learning-based models can detect patterns combining both levels, especially under a supervised approach, achieving an F1-score of 89.76\% with Random Forest and 64.09\% with LSTM-Autoencoder. Although these results indicate that the proposed models can support CSA by improving the observation and interpretation of anomalous behavior, the subtle differences between normal and attack samples highlight the need for richer discriminative features.

\end{abstract}
\section{Introduction}

In contemporary battlefields, the success of military operations depends on the integration of traditional kinetic domains, namely land, sea, air, and space, with non-kinetic operational areas: cyberspace, the Electromagnetic Spectrum (EMS), and Information Operations (IO)~\cite{Sdrakas2025,FM3-12}. Since the EMS intersects with many operational domains, including both kinetic domains and cyberspace, its control by military forces is essential for mission success. Moreover, the growing convergence between cyberspace and the electromagnetic domain has increased the need to coordinate cyber and electromagnetic capabilities to gain operational advantage. In this context, several countries have focused on the development of Cyber-Electromagnetic Activities (CEMA)~\cite{PompiliVitello_CEMA,Sdrakas2025}, which aims to coordinate offensive, defensive, informational, and enabling activities across cyberspace and the EMS. This coordination is also relevant for CSA, where heterogeneous cyber and electromagnetic evidence must be observed and interpreted to support operational understanding and decision-making.

This convergence creates the need for monitoring approaches capable of correlating evidence and characterizing events from both EMS and cyberspace perspectives. From a CSA perspective, this capability is especially relevant to the Observe and Orient phases of the OODA loop, since situational awareness (SA) depends on collecting heterogeneous indicators, interpreting their meaning, and transforming them into useful information for subsequent decision-making. Although models that jointly exploit signal-level and traffic-level information can contribute to the early identification of malicious or abnormal behavior in CEMA-related scenarios, most existing anomaly detection approaches focus on only one of these perspectives.

Within this context, this work proposes the development and evaluation of anomaly detection models for a CEMA scenario that combines characteristics of both the electromagnetic environment and cyberspace. Using the Zigbee-based ZBDS2023 dataset, which includes physical-level features, traffic-level features, and labeled replay and flooding attacks, this study analyzes two approaches for detecting anomalous behavior in temporal sequences. The first proposal is a supervised model based on Random Forest that uses statistical features computed over sequences of frames to capture temporal properties and predict whether an attack is present. This model serves as a first approach due to its simplicity. The second model is an unsupervised LSTM-Autoencoder, which learns temporal patterns from sequences and, after being trained only on normal data, determines whether a sequence is anomalous based on its reconstruction error, being one of the most used models in the literature. The results show that these approaches can detect temporal patterns combining both levels, providing useful support for CSA by improving the observation and orientation of anomalous cyber-electromagnetic behavior.

\section{Related work}
\label{sec:related_work}

Anomaly detection in electromagnetic and networked communication environments has traditionally been addressed using a single source of information. Existing approaches usually rely either on physical-layer signal features (e.g., I/Q samples, spectral power, or spectrograms) or on network-traffic features (e.g., packet headers, flow statistics, or frame metadata). As a result, there is a lack of works addressing the joint exploitation of physical-layer and traffic-level features to correlate channel events with communication behavior. For this reason, this section reviews anomaly detection methods that use physical signal features and traffic features separately. Most of the reviewed works address anomaly detection through reconstruction-based approaches. Finally, in these methods, the model learns the normal behavior of the monitored system and detects anomalies when the reconstruction error exceeds the expected range.

Focusing first on network anomalies, Casajus-Setién et al.~\cite{casajus2023_iiot_transformer} proposed an anomaly-based network intrusion detection system for industrial internet of things networks using a transformer model. Their method analyzed industrial network traffic in real time by processing sequences of IP flows. Each flow is described by aggregated statistics derived from the packets exchanged between the same source and destination IP addresses, ports, and protocol. These flow sequences were used as input to a transformer encoder-decoder model. The model was trained to reconstruct normal traffic windows, and anomaly detection was performed by computing the reconstruction loss, based on the mean squared error (MSE) between the original and reconstructed flow sequences. The method was evaluated on the WUSTL-IIoT-2021 dataset and achieved 94.31\% F1-score and 97.44\% AUC for Denial of Service (DoS) attack detection.

Similarly, Kummerow et al.~\cite{kummerow2024_transformer_explainable} proposed a transformer-based network traffic autoencoder for unsupervised anomaly detection and explanation in network packet sequences. Their model learned representations at two levels: a packet-wise level, where discrete and continuous packet features are encoded, and a temporal level, where a transformer captures dependencies across packet sequences. The anomaly score was computed from reconstruction errors: categorical accuracy was used for discrete features, and MSE was used for continuous features. The best configuration achieved 95.36\% F1-score and 95.47\% binary accuracy.

Park et al.~\cite{park2025_unsupervised_network_packets} proposed an unsupervised Intrusion Detection System (IDS) based on Long Short-Term Memory (LSTM) autoencoders, focusing specifically on a Convolutional Neural Network Bidirectional LSTM Autoencoder (CNN-BiLSTM-AE). The objective was to address the limitations of signature-based IDSs when detecting previously unseen attacks. The model was trained only on normal traffic from the CICIDS2018 dataset, after preprocessing and feature selection. Detection was performed by computing the reconstruction loss during inference and applying a predefined threshold to distinguish between normal and anomalous traffic. The model obtained 98.1\% accuracy and 98.3\% F1-score. The same model was also evaluated on the UNSW-NB15 dataset, reporting 97.7\% accuracy and 97.8\% F1-score.

In the electromagnetic spectrum domain, Tian et al.~\cite{tian2022_unsupervised_spectrum_anomaly_unauthorized} addressed unsupervised spectrum anomaly detection in unauthorized frequency bands, where spectrum composition is complex and anomaly patterns are not known in advance. They proposed a Variational Autoencoder (VAE)-based approach operating on spectrograms generated from I/Q data. The authors defined a percentile-based anomaly metric, named PER score, which focuses on the worst reconstructed regions of the spectrogram instead of averaging the reconstruction error over the entire input. Using a Conv-VAE model, the PER score achieved AUC values between 96.71\% and 98.91\% across different anomaly types, including GMSK, CHIRP, QPSK, and 16QAM signals.

Wang et al.~\cite{wang2025_adversarial_autoencoder_anomalous_radio} also focused on anomalous radio signal detection using time-frequency representations, proposing a Convolutional Neural Network Adversarial Autoencoder (CNN-AAE). Anomalies were detected using reconstruction error, with MSE permitted quantifying the discrepancy between the original and reconstructed time-frequency matrices. The model achieved accuracy values between 93.14\% and 96.71\% across the three datasets used, with F1-scores ranging from 91.2\% to 94.1\%.

A complementary line of work was presented by Lourme et al.~\cite{lourme2023_zbds2023}, who introduced ZBDS2023~\cite{lourme_hauspie2023_zigbee_dataset_cristal}. This dataset provides both MAC-layer frame data and physical-layer RSSI values and includes benign periods and labeled attacks. As a baseline, the authors evaluated a simple RSSI threshold-based IDS for spoofing detection, obtaining 91.7\% accuracy, 31.2\% precision, and 83.3\% recall for a single-device use case.

Table~\ref{table:resumen_literatura_trafico} summarizes the reviewed approaches. Most existing methods rely on reconstruction-based detection and analyze either traffic-level data or physical-layer signal representations. The research from Lourme et al.~\cite{lourme2023_zbds2023} constitutes a relevant exception, as they provide both MAC-layer Zigbee data and RSSI measurements, although their IDS is a simple threshold-based baseline. This gap motivates approaches that jointly exploit signal-level information and communication behavior.

\begin{table}[ht!]
\centering
\caption{Summary of the literature on anomaly detection in network traffic and electromagnetic spectrum.}
\label{table:resumen_literatura_trafico}
\resizebox{\columnwidth}{!}{%
\begin{tabular}{
>{\centering\arraybackslash}m{0.8cm}
>{\raggedright\arraybackslash}m{2.8cm}
>{\raggedright\arraybackslash}m{3.0cm}
>{\raggedright\arraybackslash}m{2.6cm}
>{\raggedright\arraybackslash}m{3.6cm}}
\hline
\textbf{Ref.} & \textbf{Input} & \textbf{Model} & \textbf{Anomaly score} & \textbf{Evaluation} \\
\hline
\hline
\cite{casajus2023_iiot_transformer} 
& Flow sequences
& Transformer encoder-decoder 
& MSE 
& 94.31\% F1-score; 97.44\% AUC\\
\hline
\cite{kummerow2024_transformer_explainable} 
& Packet sequences with discrete and continuous features
& Transformer-based network traffic autoencoder 
& Categorical accuracy + MSE 
& Best: 95.36\% F1-score, 95.47\% accuracy\newline Worst: 74.02\% F1-score, 79.42\% accuracy\\
\hline
\cite{park2025_unsupervised_network_packets} 
& Preprocessed traffic features 
& CNN-BiLSTM-AE 
& Reconstruction loss + threshold
& CICIDS2018: 98.3\% F1-score, 98.1\% accuracy\newline UNSW-NB15: 97.8\% F1-score, 97.7\% accuracy\\
\hline
\cite{tian2022_unsupervised_spectrum_anomaly_unauthorized} 
& Spectrograms
& Variational Autoencoder
& PER score 
& Best: 98.91\% AUC\newline Worst: 96.71\% AUC \\
\hline
\cite{wang2025_adversarial_autoencoder_anomalous_radio} 
& Time-frequency matrices
& CNN-AAE 
& MSE + threshold 
& Best: 94.1\% F1-score, 96.71\% accuracy\newline Worst: 91.2\% F1-score, 93.14\% accuracy\\
\hline
\cite{lourme2023_zbds2023} 
& Zigbee MAC-layer frames + RSSI
& RSSI threshold-based baseline IDS
& RSSI moving average + fixed thresholds
& 91.7\% accuracy, 31.2\% precision, 83.3\% recall
\\
\hline
\end{tabular}%
}
\end{table}

This work also searched for public datasets combining electromagnetic signal information and network-traffic features. The objective was to identify datasets suitable for training Deep Learning (DL) models able to detect anomalies in electromagnetic spectrum signals and to support EMS analysis in CEMA-related scenarios. Table~\ref{table:datasets_rf_trafico} summarizes the identified datasets.

Several datasets contain a limited number of samples or provide only partial coverage of the monitored environment, which may hinder the learning of robust patterns and increase the risk of overfitting. This group includes the following real datasets: \cite{duque2021sdr4iot_ble_zigbee_rf}, a Bluetooth and Zigbee dataset in the 2.4 GHz band with I/Q samples and BLE/Zigbee metadata; \cite{karoliny2022insectt_ble_channel_sniff}, a Bluetooth dataset in the 2.4 GHz band with I/Q samples and BLE channel metadata; and \cite{strohmayer_kampel2024_wallhack18k}, a Wi-Fi dataset in the 2.4 GHz band with spectrograms and raw Wi-Fi packet time series. In addition, \cite{estevez2025sigmf_pcap_annotations_example} provides synthetic data at 400 MHz with I/Q samples and PCAP annotations. Although these resources may be useful for specific studies, none of them include labeled anomalies, which limits their applicability for supervised or evaluation-oriented anomaly detection.

Datasets based on LoRaWAN technologies, such as \cite{bhatia2020loed,povalac_kral2023_lorawan_traffic_analysis}, are especially relevant because they provide physical-layer indicators, such as RSSI or SNR, together with packet-related information. Both are datasets operating in the 868 MHz band: \cite{bhatia2020loed} provides LoRaWAN packet and PHY metadata, while \cite{povalac_kral2023_lorawan_traffic_analysis} combines LoRaWAN PHY and packet metadata. However, LoRaWAN traffic is typically sparse, which introduces significant temporal gaps between consecutive samples that can complicate the training of sequential models to capture short-term dynamics or rapid variations in the radio channel.

An intermediate case is \cite{elmaghbub_hamdaoui_stable_wifi_rf_fingerprints_dataset}, a real Wi-Fi dataset in the 2.4 GHz band that includes I/Q samples and Wi-Fi packet metadata. However, the dataset was collected using a constant header and a payload of 0 bytes. Although it is useful for RF fingerprinting, its traffic-level variability is limited, reducing its suitability for analyzing realistic communication patterns associated with incidents or attacks. Another relevant resource is \cite{hackveda_darpa_sc2_repo}, a synthetic dataset from the DARPA Spectrum Collaboration Challenge at 100 MHz, which contains a large volume of data and includes PSD, RF and frame-level metadata. However, the lack of sufficient metadata to contextualize each record makes the dataset difficult to interpret and reuse for anomaly detection studies.

Finally, \cite{bravenec_etal2023_uji_probes_supplementary,lourme_hauspie2023_zigbee_dataset_cristal} provide real data in the 2.4 GHz band with both physical-layer and traffic-level information. The former is a Wi-Fi dataset with Radiotap-based metadata, but without labeled anomalies. In contrast, \cite{lourme_hauspie2023_zigbee_dataset_cristal} is a Zigbee dataset that includes MAC-layer metadata and RSSI values captured from multiple locations, as well as benign periods and labeled attack scenarios. Although the dataset focuses on Zigbee, this technology is representative of short-range, low-power wireless communications commonly used in sensor-rich environments, making it suitable for approximating scenarios where spectrum monitoring and early detection of interference or attacks are relevant.

This review shows that there is a lack of publicly available datasets that jointly provide physical-layer radio-frequency information, network traffic traces or metadata, and representative attack scenarios with consistent annotations. Considering these limitations and the available data volume, the Zigbee dataset in \cite{lourme_hauspie2023_zigbee_dataset_cristal} was selected as the reference dataset for the development of this work.

\begin{table}[ht!]
\centering
\caption{Datasets with EMS information and network-traffic features.}
\label{table:datasets_rf_trafico}
\resizebox{\columnwidth}{!}{%
\begin{tabular}{
>{\centering\arraybackslash}m{0.8cm}
>{\raggedright\arraybackslash}m{1.9cm}
>{\centering\arraybackslash}m{1.8cm}
>{\raggedright\arraybackslash}m{5.2cm}
>{\centering\arraybackslash}m{1.0cm}
>{\centering\arraybackslash}m{1.8cm}}
\hline
\textbf{Ref.} & \textbf{Technology} & \textbf{Frequency} & \textbf{Features} & \textbf{Data} & \textbf{Anomalies} \\
\hline
\hline
\cite{bhatia2020loed} 
& LoRaWAN 
& 868 MHz 
& LoRaWAN packet and PHY metadata
& R
& - \\
\hline
\cite{bravenec_etal2023_uji_probes_supplementary} 
& Wi-Fi 
& 2.4 GHz 
& Radiotap-based Wi-Fi metadata
& R
& - \\
\hline
\cite{duque2021sdr4iot_ble_zigbee_rf} 
& Bluetooth and Zigbee 
& 2.4 GHz 
& I/Q samples and BLE/Zigbee metadata
& R
& - \\
\hline
\cite{elmaghbub_hamdaoui_stable_wifi_rf_fingerprints_dataset} 
& Wi-Fi 
& 2.4 GHz 
& I/Q samples and Wi-Fi packet metadata
& R
& - \\
\hline
\cite{estevez2025sigmf_pcap_annotations_example} 
& Not specified 
& 400 MHz 
& I/Q samples and PCAP annotations 
& S 
& - \\
\hline
\cite{hackveda_darpa_sc2_repo} 
& Not specified 
& 100 MHz 
& PSD, RF and frame-level metadata
& S
& - \\
\hline 
\cite{karoliny2022insectt_ble_channel_sniff} 
& Bluetooth 
& 2.4 GHz 
& I/Q samples and BLE channel metadata 
& R
& - \\
\hline
\cite{lourme_hauspie2023_zigbee_dataset_cristal} 
& Zigbee 
& 2.4 GHz 
& MAC-layer metadata and RSSI 
& R
& \checkmark \\
\hline
\cite{povalac_kral2023_lorawan_traffic_analysis} 
& LoRaWAN 
& 868 MHz 
& LoRaWAN PHY and packet metadata
& R 
& - \\
\hline
\cite{strohmayer_kampel2024_wallhack18k} 
& Wi-Fi 
& 2.4 GHz
& Spectrograms and raw Wi-Fi packet time series 
& R 
& - \\
\hline
\end{tabular}%
}
\begin{tablenotes}
  \small
  \item R: Real data; S: Synthetic data.
\end{tablenotes}
\end{table}

\section{Design and Implementation of the Proposed Solution}
\label{sec:design}

\begin{figure}[ht!]
    \centering
    \includegraphics[width=1\linewidth]{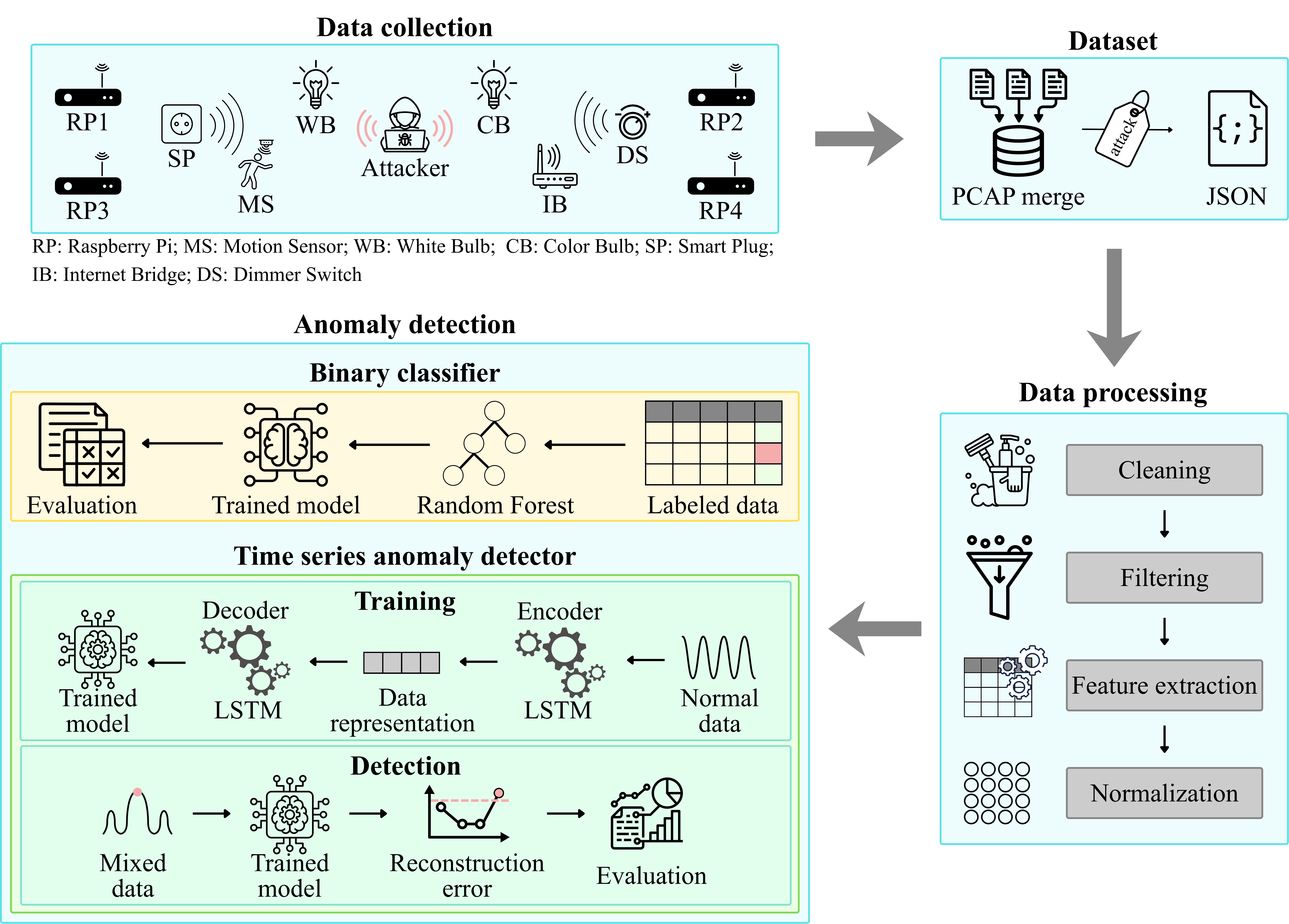}
    \caption{Overview of the proposed scenario for CEMA anomaly detection.}
    \label{fig:esquema_diseño}
\end{figure}

The implemented solution follows the workflow shown in Figure~\ref{fig:esquema_diseño}. First, data are collected in a real scenario composed of a mesh network of Zigbee devices. Then, the captured traffic is processed for cleaning, feature extraction, labeling, and normalization. Based on the processed data, two anomaly detection perspectives are explored. On the one hand, a supervised binary classifier based on Random Forest determines whether a group of frames corresponds to normal behavior or to an attack. On the other hand, an unsupervised time-series detector based on an LSTM-Autoencoder learns the normal behavior of the system from temporal sequences and detects deviations through the reconstruction error.

This workflow was implemented in Python using Scikit-learn and Keras. Scikit-learn was used for processing, normalization, dataset splitting, evaluation metrics, and the Random Forest classifier, while Keras was used to build and train the LSTM-Autoencoder model. The source code is publicly available in the GitHub repository~\cite{CEMA_github}.

\subsection{Dataset: ZBDS2023}

For the experimental development of this work, this work used the ZBDS2023 dataset~\cite{lourme_hauspie2023_zigbee_dataset_cristal}. This is the same dataset introduced and analyzed by Lourme et al.~\cite{lourme2023_zbds2023}, as discussed in the previous section. This dataset was designed to support the development and evaluation of intrusion detection systems (IDSs) in Zigbee-based IoT environments. It was collected in an inhabited smart home equipped with 10 Philips Hue devices. The dataset includes both normal operation periods and periods with attacks.

The capture setup consists of four passive probes distributed across different locations in the house. Each probe is built using a Raspberry Pi and a CC2531 module. This configuration allows a single frame transmitted by a Zigbee device to be captured by one to four probes, depending on the coverage conditions. As a result, the dataset provides redundancy at the MAC layer and, additionally, RSSI measurements from different spatial locations. The captures are stored in PCAP format and organized by probe and time interval. Additionally, the ZBDS2023 dataset includes 50 labeled attacks distributed across multiple sessions. In total, five attack types are considered, labeled from A to E attacks and detailed in~\cite{lourme_hauspie2023_zigbee_dataset_cristal}, including four different flooding strategies and one replay attack.

To facilitate the preprocessing stage, the dataset authors provide code for generating JSON files from the original PCAP captures. In this work, that code was adapted to include additional fields in the generated JSON records. As a result, each processed record contains the attributes summarized in Table~\ref{table:features_dataset}. Together, these attributes allow the construction of device-level time series and the analysis of traffic behavior along with physical-layer information.

\begin{table}[ht!]
\centering
\caption{Features extracted from the original dataset.}
\label{table:features_dataset}
\resizebox{\columnwidth}{!}{%
\begin{tabular}{
>{\raggedright\arraybackslash}m{3.0cm}
>{\raggedright\arraybackslash}m{8.6cm}}
\hline
\textbf{Feature} & \textbf{Description} \\
\hline
\hline
\texttt{frame\_index} & Frame index derived from the Wireshark column index. \\
\hline
\texttt{src16} & 16-bit MAC short address of the sender. \\
\hline
\texttt{length} & Frame length in bytes. \\
\hline
\texttt{time\_relative\_us} & Time elapsed since the first frame, expressed in $\mu$s. \\
\hline
\texttt{rssi} & RSSI value associated with the captured frame. \\
\hline
\texttt{epoch\_us} & Frame reception time in Unix epoch format, expressed in $\mu$s. \\
\hline
\texttt{dst16} & 16-bit MAC short address of the receiver. \\
\hline
\texttt{attack} & Attack label associated with the frame. Possible values are: \texttt{NO\_ATTACK}, \texttt{A}, \texttt{B}, \texttt{C}, \texttt{D}, and \texttt{E}. \\
\hline
\texttt{dev} & Reference of the device transmitting the frame, described in~\cite{lourme2023_zbds2023}. \\
\hline
\texttt{rpi} & Identifier of the Raspberry Pi probe that captured the frame. \\
\hline
\end{tabular}%
}
\end{table}

The preprocessing code used to obtain the JSON files from the PCAP captures extracts the variables described above and applies filtering to retain only well-formed and valid frames. As a result, all generated records contain complete information, without null values or empty fields. After preprocessing, approximately 8\,500\,000 valid samples are obtained per Raspberry Pi, resulting in nearly 34\,000\,000 rows with 10 columns.

Table~\ref{table:etiquetas_attack} shows the number of samples labeled with each attack type and with the \texttt{NO\_ATTACK} label. In total, 33\,829\,336 rows correspond to normal operation, whereas 55\,638 rows are associated with attack contexts. This evidences a strong class imbalance, since attack samples represent only 0.16\% of the complete dataset. Furthermore, the dataset documentation indicates that the attack sessions took place between July 3rd and July 9th. Therefore, when the analysis is restricted to this time interval, the number of non-attack rows is reduced to 17\,721\,706, increasing the attack ratio to 0.31\%.

\begin{table}[ht!]
\centering
\caption{Number of samples associated with each attack label.}
\label{table:etiquetas_attack}
\resizebox{0.55\columnwidth}{!}{%
\begin{tabular}{
>{\centering\arraybackslash}m{3.0cm}
>{\centering\arraybackslash}m{3.5cm}}
\hline
\textbf{Label} & \textbf{Number of samples} \\
\hline
\hline
\texttt{NO\_ATTACK} & 33\,829\,336 \\
\texttt{A} & 6\,755 \\
\texttt{B} & 6\,268 \\
\texttt{C} & 2\,919 \\
\texttt{D} & 20\,599 \\
\texttt{E} & 19\,097 \\
\hline
\end{tabular}%
}
\end{table}

\subsection{Data processing}

To use the data effectively, a preliminary processing stage was applied, including cleaning, filtering, label transformation, and feature engineering. First, frames transmitted or received by unknown devices were discarded. Regarding attack labeling, the decision was made not to distinguish between specific attack types, but rather to formulate the task as a binary classification problem. Therefore, the original labels were mapped to 0 for normal samples and 1 for attack samples.

For model training, several metadata fields were removed because they do not provide relevant predictive information or could make the model overly dependent on the specific capture setup. These discarded fields include the Raspberry Pi that captured the frame (\texttt{rpi}), the frame index (\texttt{frame\_index}), and the relative time with respect to the first frame (\texttt{time\_relative\_us}). Variables directly associated with device identifiers or device types were also excluded. This design choice aims to obtain a more general model, less dependent on the specific devices and probes present in the experimental environment. Additionally, to incorporate temporal information, two derived variables were defined: \texttt{delta\_t}, which represents the time difference between consecutive samples emitted by the same device and captured by the same Raspberry Pi, and \texttt{delta\_l}, which represents the length difference between those consecutive samples. These variables introduce part of the temporal behavior of the traffic into models that are not originally designed to process temporal sequences. They also facilitate feature extraction for models that can explicitly exploit sequential information.

\subsection{Anomaly detection}

This section describes the process followed to implement two models aimed at detecting anomalies and attacks in the previously described dataset.

\subsubsection{Binary anomaly detection.}

This first approach developed a supervised binary classification model based on Random Forest. The objective is to determine, for a fixed-size window of samples, based on the statistical features of each window, whether an attack occurred (1) or not (0).

To preserve temporal information, each window is constructed from consecutive messages emitted by the same device and captured by the same Raspberry Pi. Once the window is defined, the following statistical measures are computed for each feature: mean, median, minimum, maximum, 25th percentile, 75th percentile, and interquartile range (IQR). The resulting vector, composed of these aggregated statistics, is used as input during both training and prediction phases.

The model requires samples labeled as normal (0) and attack (1). Therefore, to reduce the imbalance between these two classes, this work only used the time interval between July 3rd and July 9th, which corresponds to the attack sessions. As a result, the dataset used for this approach contains 17\,721\,371 normal samples and 55\,638 attack samples.

\paragraph{Evaluation Metrics.} 

Because the dataset presents a marked imbalance between normal and attack samples, metrics such as F1-score are more appropriate than accuracy, since they better reflect the balance between detecting attacks and avoiding false alarms. For this reason, the primary metric used to select the best model was F1-score, although precision and recall are also employed to obtain a more detailed evaluation.

    
    

\paragraph{Training Procedure.}
\label{sec:RF_training}

Given the large volume of available data, a standard partitioning strategy was used while preserving the temporal order of the samples. Specifically, the first 80\% of the data, ordered by date, was used for training, and the subsequent 20\% was reserved for testing. Hyperparameter tuning and five-fold cross-validation were performed on the training set, while the test set was kept isolated until the final evaluation.

To properly train the model, it was necessary to determine which combination of hyperparameters provided the best results. After an initial search, a set of candidate values was selected for each of the hyperparameters. The most promising values are summarized in Table~\ref{table:RF_hyperparam_values}. Using this hyperparameter search space and five-fold cross-validation, the configuration that optimized the model according to the selected evaluation metric was identified and subsequently used for final training.

\begin{table}[ht!]
\centering
\caption{Values used in the Random Forest hyperparameter search.}
\label{table:RF_hyperparam_values}
\resizebox{0.5\columnwidth}{!}{%
\begin{tabular}{
>{\raggedright\arraybackslash}m{3.0cm}
>{\raggedright\arraybackslash}m{3.0cm}}
\hline
\textbf{Hyperparameter} & \textbf{Values} \\
\hline
\hline
\texttt{n\_estimators} & 200, 400, 800 \\
\hline
\texttt{max\_features} & \texttt{None}, \texttt{sqrt} \\
\hline
\texttt{max\_depth} & \texttt{None}, 80, 120 \\
\hline
\texttt{min\_samples\_split} & 20, 40, 50 \\
\hline
\texttt{min\_samples\_leaf} & 1 \\
\hline
\texttt{bootstrap} & \texttt{True} \\
\hline
\texttt{criterion} & \texttt{gini}, \texttt{entropy} \\
\hline
\texttt{class\_weight} & \texttt{None} \\
\hline
\end{tabular}%
}
\end{table}

\subsubsection{Detection of anomalies in time series.}

To address anomaly detection from a temporal perspective, an unsupervised LSTM-Autoencoder was implemented. The model was trained only with attack-free traffic and learned to reconstruct normal sequences. During inference, sequences with reconstruction errors above a selected threshold were classified as anomalous.

\paragraph{Model Architecture.}
The LSTM-Autoencoder architecture is organized into two main blocks: an encoder and a decoder. The encoder consists of LSTM layers interleaved with \textit{LayerNormalization}, which compress the input sequence into a temporal representation. A \textit{RepeatVector} layer adapts this representation to the original sequence length, and the decoder reconstructs the input sequence. Finally, a \textit{TimeDistributed} dense layer generates the reconstructed variables at each time step. The model uses the Adam optimizer and \textit{Early Stopping} to reduce overfitting.

\paragraph{Data preprocessing.}
Before training, the data are reshaped into three-dimensional inputs with the following structure: (number of samples, timesteps, number of features). Sequences are generated using a window of length \texttt{seq\_length} and an offset \texttt{stride}. To preserve temporal consistency, each sequence is built from consecutive samples emitted by the same device and captured by the same Raspberry Pi. For evaluation, a test sequence is labeled as anomalous if, at least, 20\% of its samples are associated with an attack; otherwise, it is labeled as normal.

\paragraph{Evaluation metric and anomaly threshold.}

The evaluation prioritizes high recall, since in cybersecurity contexts it is especially important to minimize undetected attacks, even at the cost of some false positives. However, precision and F1-score are also considered to assess the trade-off between detection capability and false alarm rate.

The LSTM-Autoencoder does not directly output a class label. Instead, it reconstructs the input sequence, so anomaly detection requires measuring the difference between the original and reconstructed sequences. For this purpose, MSE is used:

\[
\text{MSE}(X,\hat{X}) = \frac{1}{T \cdot F}\sum_{t=1}^{T}\sum_{f=1}^{F} \left(x_{t,f} - \hat{x}_{t,f}\right)^2
\]

where \(T\) is the sequence length, \(F\) is the number of features, \(X\) is the original sequence, and \(\hat{X}\) is the reconstructed sequence.

\begin{figure}[ht!]
    \centering
    \includegraphics[width=\linewidth]{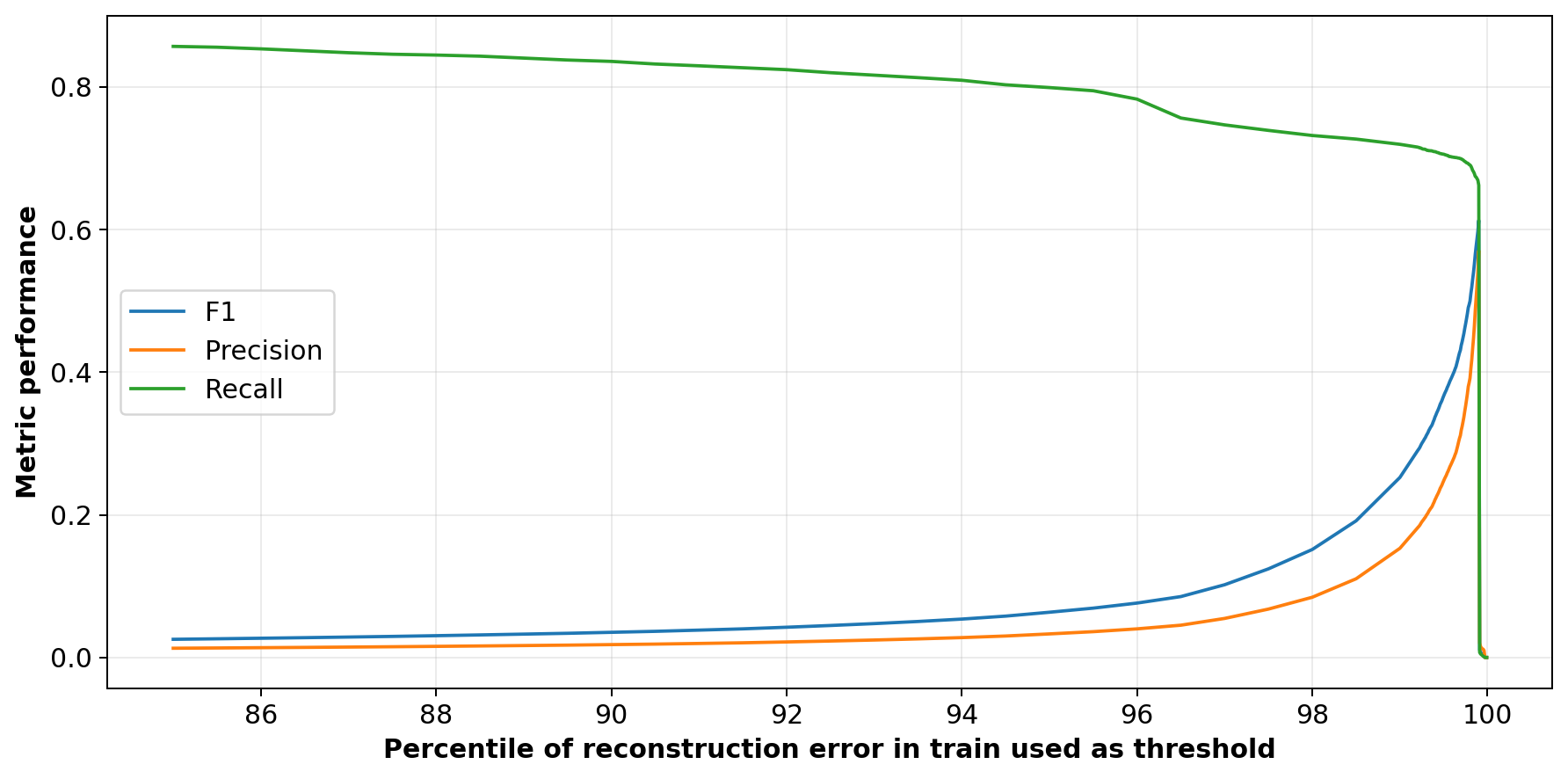}
    \caption{Evolution of the F1-score, precision, and recall as a function of the percentile used to define the anomaly threshold. The threshold is obtained from the reconstruction error distribution of normal sequences and evaluated on a mixed validation set with normal and attack traffic. The configuration shown corresponds to a model with four LSTM layers, \texttt{seq\_length} = 8, \texttt{stride} = 4, \texttt{units} = 128, and \texttt{middle\_units} = 64.}
    \label{fig:evolucion_percentiles}
\end{figure}

The anomaly threshold is selected as a percentile of the reconstruction error distribution under normal conditions and is tuned on a mixed validation set containing normal and under attack sequences. Once selected, the threshold remains fixed during testing. Figure~\ref{fig:evolucion_percentiles} illustrates the evolution of the F1-score, precision, and recall metrics as the percentile used to define the threshold increases.

\paragraph{Training procedure.}
\label{sec:LSTM_training}

The training set was restricted to samples transmitted between June 30th and July 3rd, which are completely free of attacks. In contrast, the interval between July 3rd and July 9th, where normal and anomalous sequences coexist, was used for threshold selection and final evaluation.

The attack-free subset was divided into 80\% for training and 20\% for validation. The training subset was used to adjust the model weights, while the validation subset was utilized to monitor the reconstruction loss. The subset containing both normal and malicious traffic was also divided into two parts: 50\% for validation and 50\% for testing. The mixed validation subset was used to select both the anomaly threshold and the hyperparameter configuration.

Some hyperparameters were kept fixed across all experiments using standard values, while the remaining ones were explored experimentally. The fixed values are highlighted with a gray background in Table~\ref{table:LSTM_hyperparameters}. Starting from this baseline, a hyperparameter exploration was performed to identify the architecture and parameterization that maximized detection capability on the validation data. An initial manual analysis showed that the best-performing architecture consisted of two LSTM layers in the encoder and two LSTM layers in the decoder. The most promising values for the remaining hyperparameters are shown in Table~\ref{table:LSTM_hyperparam_values}.

\begin{table}[ht!]
\centering
\caption{Values used in the LSTM-Autoencoder hyperparameter search.}
\label{table:LSTM_hyperparam_values}
\resizebox{0.7\columnwidth}{!}{%
\begin{tabular}{
>{\raggedright\arraybackslash}m{3.0cm}
>{\raggedright\arraybackslash}m{6.0cm}}
\hline
\textbf{Hyperparameter} & \textbf{Values} \\
\hline
\hline
\texttt{sequence length} & 8, 16 \\
\hline
\texttt{stride} & \texttt{sequence length}/2, \texttt{sequence length} \\
\hline
\texttt{units} & 128, 256 \\
\hline
\texttt{middle\_units} & 64, 96, 128 \\
\hline
\end{tabular}%
}
\end{table}

Within the list of hyperparameters, \texttt{units} refers to the number of hidden units in the first LSTM layer of the encoder and the last LSTM layer of the decoder, while \texttt{middle\_units} refers to the number of hidden units in the internal LSTM layers. From this search, the most suitable hyperparameter combination for the proposed time-series anomaly detector was selected.

\section{Results}
\label{sec:results}

This section summarizes the results obtained by the supervised Random Forest model and the unsupervised LSTM-Autoencoder model.

\subsection{Random Forest-Based Binary Detection}

The most suitable window configuration, defined by \texttt{seq\_length} and \texttt{stride}, can be determined by analyzing the length of the attack sequences. When consecutive attack samples are grouped by device and Raspberry Pi, following the same criterion used to construct the input windows, most attack sequences are concentrated within a relatively short range, approximately between 1 and 6 frames, as shown in Table~\ref{table:seq_length_total}. This supports the idea that excessively long sequences may dilute the attack signal, whereas shorter or intermediate windows can better capture local variations associated with attacks. For this reason, sequences of length 8 were selected. In addition, preliminary experiments with a simple frame-based Random Forest model were conducted to determine the most suitable stride value, resulting in a stride of 8. These results are not presented in this paper for simplicity. 

\begin{table}[ht!]
\centering
\caption{Partial distribution of attack sequences according to their length. Entries in bold indicate the most common sequence lengths.}
\label{table:seq_length_total}
\resizebox{\columnwidth}{!}{%
\begin{tabular}{
|>{\centering\arraybackslash}m{2.0cm}
>{\centering\arraybackslash}m{2.0cm}|
>{\centering\arraybackslash}m{2.0cm}
>{\centering\arraybackslash}m{2.0cm}|
>{\centering\arraybackslash}m{2.0cm}
>{\centering\arraybackslash}m{2.0cm}|
>{\centering\arraybackslash}m{2.0cm}
>{\centering\arraybackslash}m{2.0cm}|}
\hline
Sequence length & Number of sequences & Sequence length & Number of sequences & Sequence length & Number of sequences & Sequence length & Number of sequences \\
\hline
\hline
\textbf{1}  & \textbf{642} & 8  & 1  & 15 & 34 & 22 & 42 \\
\textbf{2}  & \textbf{177} & 9  & 3  & 16 & 36 & 23 & 24 \\
\textbf{3}  & \textbf{52}  & 10 & 6  & 17 & 41 & 24 & 8  \\
\textbf{4}  & \textbf{141} & 11 & 13 & 18 & 35 & 25 & 14 \\
\textbf{5}  & \textbf{91}  & 12 & 14 & 19 & 64 & 26 & 8  \\
\textbf{6}  & \textbf{101} & 13 & 25 & 20 & 43 & 27 & 16 \\
7           & 36           & 14 & 53 & 21 & 28 & 28 & 7  \\
\hline
\end{tabular}%
}
\end{table}

The best configuration obtained after hyperparameter tuning is shown in Table~\ref{table:hyperparameters_RF}. The results, summarized in Table~\ref{table:RF_results_windows}, show that this model reaches an F1-score of 89.76\% on the test set, with a precision of 97.74\% and a recall of 82.99\%. The resulting confusion matrix was $\begin{bmatrix} 745743 & 81 \\ 719 & 3508 \end{bmatrix}$, where rows correspond to the actual classes, normal and attack, and columns correspond to the predicted classes, normal and attack, respectively. The high precision value indicates that most windows classified as attacks correspond to anomalous behavior, resulting in a low false alarm rate. At the same time, the recall value shows that the model is capable of identifying most attack instances, although some anomalous windows are still misclassified as normal. Overall, the obtained F1-score reflects a strong balance between detection capability and classification reliability. This performance suggests that aggregating consecutive frames in statistical values provides a robust representation of local behavior and enables discrimination between normal and attack traffic.

\begin{table}[ht!]
\centering
\caption{Selected hyperparameters for the Random Forest model.}
\label{table:hyperparameters_RF}
\resizebox{0.4\columnwidth}{!}{%
\begin{tabular}{
>{\raggedright\arraybackslash}m{3.5cm}
>{\raggedright\arraybackslash}m{1.5cm}}
\hline
\textbf{Hyperparameter} & \textbf{Value} \\ 
\hline
\hline
\texttt{n\_estimators} & 800 \\ 
\hline
\texttt{max\_features} & \texttt{None} \\ 
\hline
\texttt{max\_depth} & 80 \\ 
\hline
\texttt{min\_samples\_split} & 20 \\ 
\hline
\texttt{min\_samples\_leaf} & 1 \\ 
\hline
\texttt{bootstrap} & \texttt{True} \\
\hline
\texttt{criterion} & \texttt{entropy} \\
\hline
\texttt{class\_weight} & \texttt{None} \\
\hline
\end{tabular}%
}
\end{table}

\begin{table}[ht!]
\centering
\caption{Results obtained on the test set with the Random Forest model.}
\label{table:RF_results_windows}
\resizebox{0.8\columnwidth}{!}{%
\begin{tabular}{
>{\centering\arraybackslash}m{2.0cm}
>{\centering\arraybackslash}m{2.0cm}
>{\centering\arraybackslash}m{2.0cm}
>{\centering\arraybackslash}m{2.0cm}
>{\centering\arraybackslash}m{2.0cm}}
\hline
\textbf{Accuracy} & \textbf{Balanced accuracy} & \textbf{Precision} & \textbf{Recall} & \textbf{F1-score}\\
\hline
\hline
99.89\% & 91.49\% & 97.74\% & 82.99\% & 89.76\%\\
\hline
\end{tabular}%
}
\end{table}

\subsection{LSTM-Autoencoder-Based Temporal Detection}

\begin{figure}[ht!]
    \centering
    \includegraphics[width=0.8\linewidth]{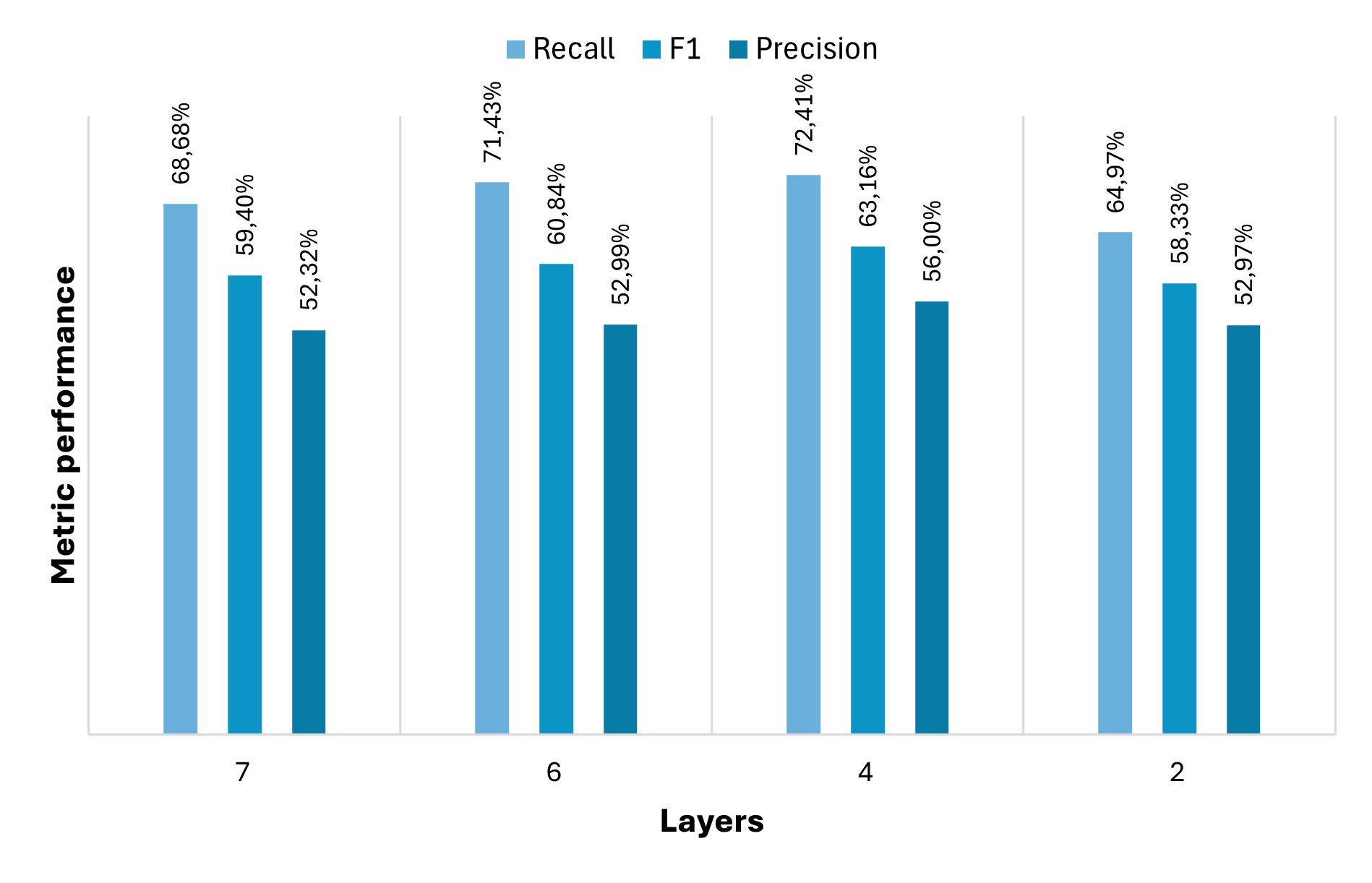}
    \caption{Comparison of the LSTM-Autoencoder performance when varying the number of layers, with \texttt{seq\_length} = 8 and \texttt{stride} = 4. The 7-layer architecture uses 192/144/96/48-96/144/192 units, the 6-layer architecture uses 256/128/64-64/128 units, the 4-layer architecture uses 128/64-64/128 units, and the 2-layer architecture uses 256-256 units.}
    \label{fig:comparativaLSTMlayers}
\end{figure}

The unsupervised temporal detector was evaluated using different architectural and sequence configurations. First, 2-, 4-, 6-, and 7-layer architectures were compared, as shown in Figure~\ref{fig:comparativaLSTMlayers}. The results indicate that increasing model complexity improves performance only up to a certain point. Based on these results, the 4-layer architecture provides the best balance between recall, precision, and F1-score.

\begin{figure}[ht!] 
    \centering 
    \begin{subfigure}[b]{0.48\textwidth} 
        \centering 
        \includegraphics[width=\textwidth]{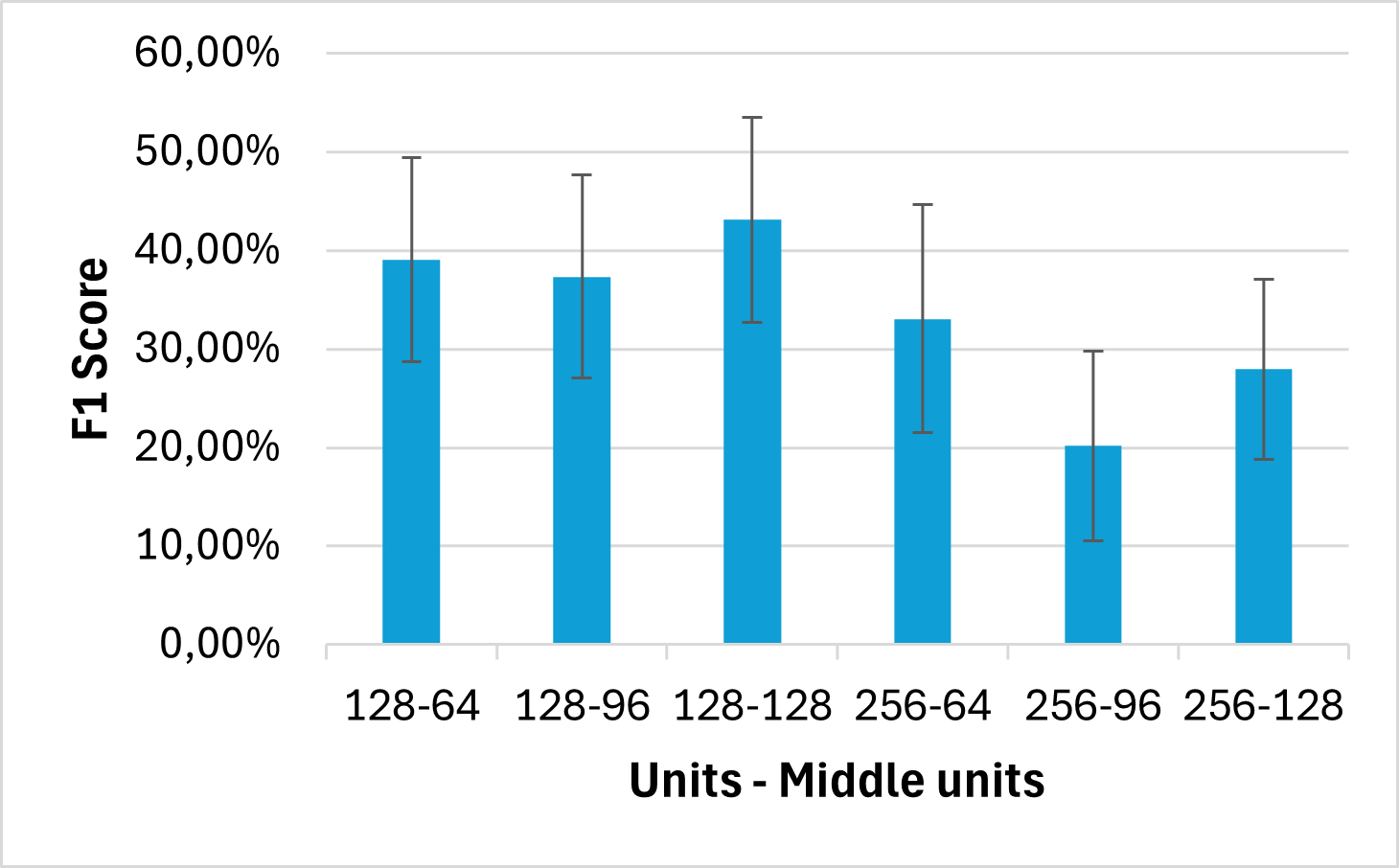} 
        \caption{\texttt{seq\_length} = 8, \texttt{stride} = 4.} 
        \label{subfig:hiperLSTM8-4} 
    \end{subfigure} 
    \hfill 
    \begin{subfigure}[b]{0.48\textwidth} 
        \centering 
        \includegraphics[width=\textwidth]{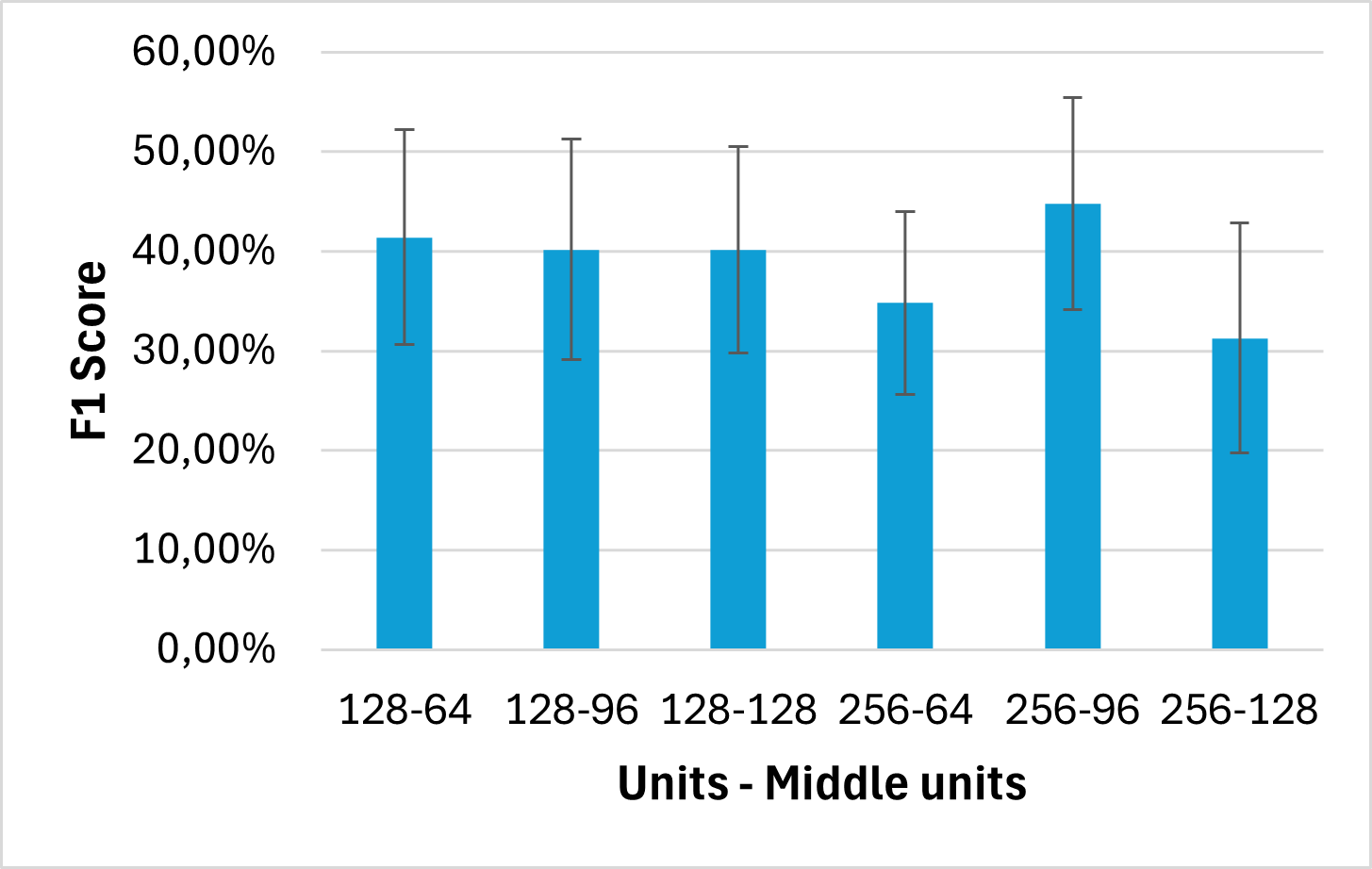} 
        \caption{\texttt{seq\_length} = 8, \texttt{stride} = 8.} 
        \label{subfig:hiperLSTM8-8} 
    \end{subfigure}
    
    \vspace{0.5cm}

    \begin{subfigure}[b]{0.48\textwidth}
        \centering
        \includegraphics[width=\textwidth]{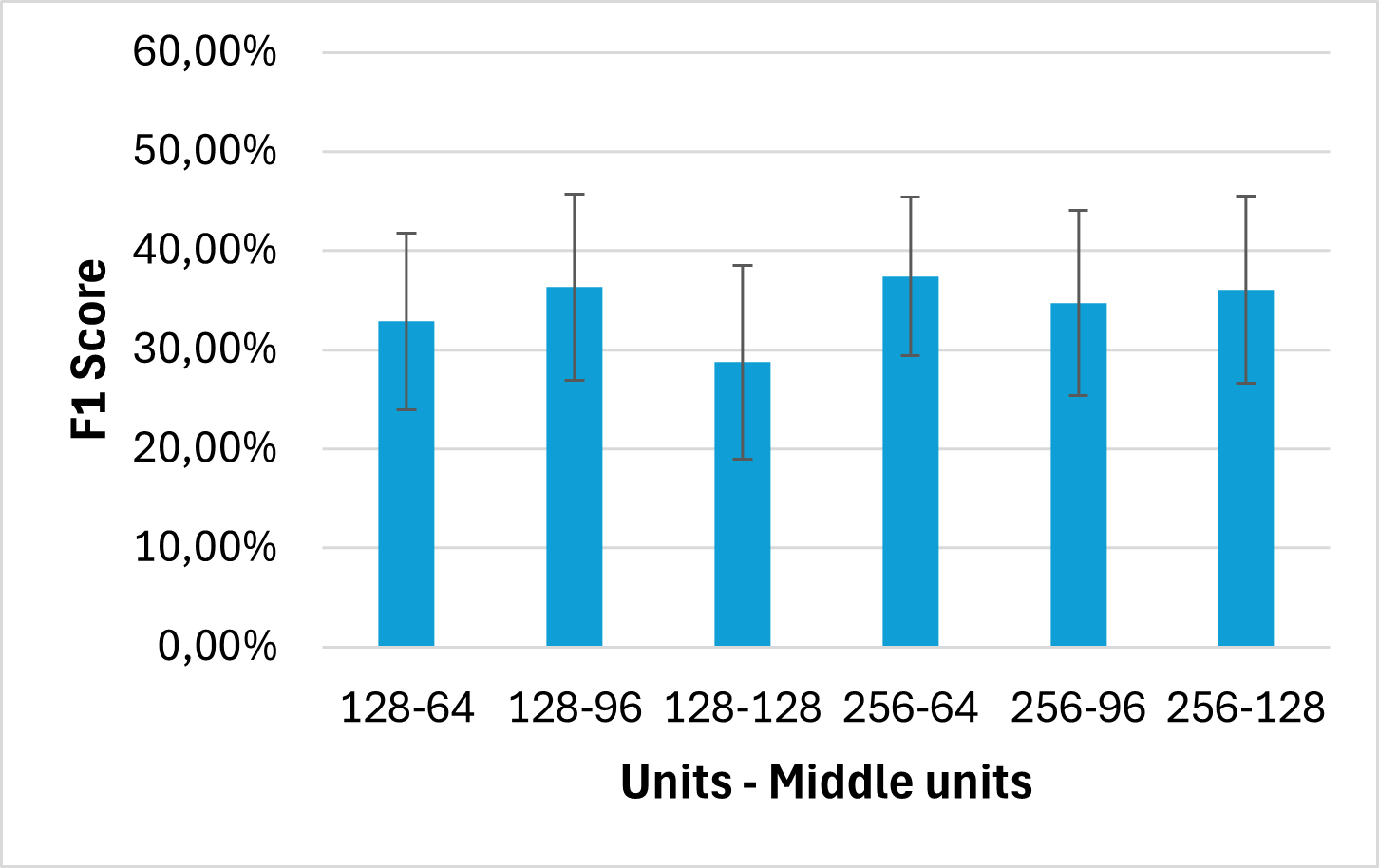}
        \caption{\texttt{seq\_length} = 16, \texttt{stride} = 8.}
        \label{subfig:hiperLSTM16-8}
    \end{subfigure}
    \hfill 
    \begin{subfigure}[b]{0.48\textwidth} 
        \centering 
        \includegraphics[width=\textwidth]{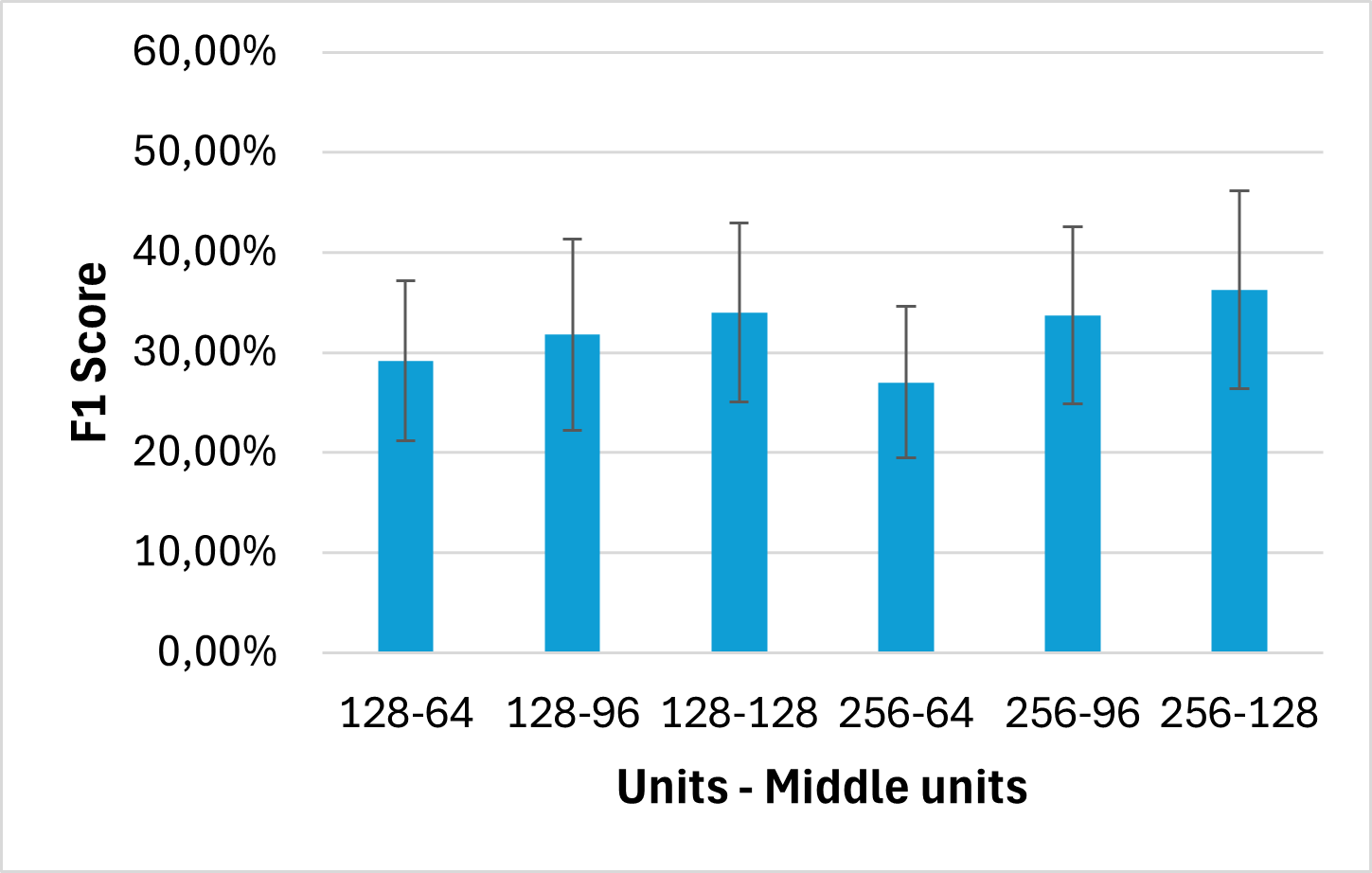} 
        \caption{\texttt{seq\_length} = 16, \texttt{stride} = 16.} 
        \label{subfig:hiperLSTM16-16} 
    \end{subfigure}
    
    \caption{Results of the hyperparameter search for the LSTM-Autoencoder model.}
    \label{fig:LSTM_hyperparam_search}
\end{figure}

These preliminary analysis narrowed the hyperparameter search space described in Section~\ref{sec:LSTM_training}. Then, a final search was performed to select the most suitable configuration, leading to the results shown in Figure~\ref{fig:LSTM_hyperparam_search}. Based on this search, the selected configuration was a 4-layer LSTM-Autoencoder with \texttt{seq\_length} = 8, \texttt{stride} = 8, \texttt{units} = 256, and \texttt{middle\_units} = 96. The complete configuration is summarized in Table~\ref{table:LSTM_hyperparameters}.

\begin{table}[ht!]
\centering
\caption{Selected hyperparameters for the \textit{LSTM-Autoencoder} model. Hyperparameters kept fixed during tuning are highlighted with a gray background.}
\label{table:LSTM_hyperparameters}
\resizebox{0.4\columnwidth}{!}{%
\begin{tabular}{
>{\raggedright\arraybackslash}m{3.5cm}
>{\raggedright\arraybackslash}m{1.5cm}}
\hline
\textbf{Hyperparameter} & \textbf{Value} \\ 
\hline
\hline
\texttt{sequence length} & 8 \\
\hline
\texttt{stride} & 8 \\
\hline
\texttt{units} & 256 \\
\hline
\texttt{middle\_units} & 96 \\
\hline
\rowcolor{gray!15}
\texttt{loss} & \texttt{mse} \\
\hline
\rowcolor{gray!15}
\texttt{activation} & \texttt{tanh} \\
\hline
\rowcolor{gray!15}
\texttt{recurrent activation} & \texttt{sigmoid} \\
\hline
\rowcolor{gray!15}
\texttt{batch size} & 64 \\
\hline
\rowcolor{gray!15}
\texttt{epochs} & 60 \\
\hline
\rowcolor{gray!15}
\texttt{learning rate} & 0.0001 \\
\hline
\rowcolor{gray!15}
\texttt{lambda} & 0.00001 \\
\hline
\rowcolor{gray!15}
\texttt{dropout rate} & 0.15 \\
\hline
\rowcolor{gray!15}
\texttt{recurrent dropout} & 0 \\
\hline
\end{tabular}%
}
\end{table}

The final results, presented in Table~\ref{table:LSTM_results}, show that the model achieves a balanced accuracy of 86.55\% and a recall of 73.36\%, indicating that it detects a relevant proportion of anomalous windows, as also reflected in the confusion matrix $\begin{bmatrix} 931085 & 2313 \\ 1109 & 3054 \end{bmatrix}$. Nevertheless, its precision of 56.90\% and F1-score of 64.09\%  are lower than the performance achieved by the supervised Random Forest model. These results are consistent with the nature of the model. Unlike Random Forest, the LSTM-Autoencoder is trained only on normal traffic and detects attacks indirectly through reconstruction error. Therefore, its performance depends on how clearly attack sequences deviate from normal behavior in the selected feature space. Although less precise, this unsupervised approach remains relevant for CSA because it can provide a warning mechanism in scenarios where labeled attack data are scarce.

\begin{table}[ht!]
\centering
\caption{Results obtained on the test set with the LSTM-Autoencoder model.}
\label{table:LSTM_results}
\resizebox{0.8\columnwidth}{!}{%
\begin{tabular}{
>{\centering\arraybackslash}m{2.0cm}
>{\centering\arraybackslash}m{2.0cm}
>{\centering\arraybackslash}m{2.0cm}
>{\centering\arraybackslash}m{2.0cm}
>{\centering\arraybackslash}m{2.0cm}}
\hline
\textbf{Accuracy} & \textbf{Balanced accuracy} & \textbf{Precision} & \textbf{Recall} & \textbf{F1-score}\\
\hline
\hline
99.63\% & 86.55\% & 56.90\% & 73.36\% & 64.09\% \\
\hline
\end{tabular}%
}
\end{table}

\section{Conclusions}
\label{sec:conclusions}
The results show that the supervised model substantially improves the accuracy and precision of the original IDS baseline proposed by the dataset authors, while maintaining a similar recall. In addition, the proposed approach is more general, since it does not rely on specific devices or fixed device-dependent RSSI profiles. From a CSA perspective, the obtained performance suggests that cyber-electromagnetic anomaly detection models can contribute to the Observe and Orient phases of the OODA loop by supporting the detection and interpretation of abnormal behavior across both the electromagnetic and cyber domains.

The best results are obtained with the window-based Random Forest model. These results suggest that aggregating consecutive frames provides a robust representation of local traffic behavior and improves the discrimination between normal and attack traffic. The unsupervised LSTM-Autoencoder also provides promising results, although its performance remains below that of the supervised Random Forest model. This behavior is consistent with the fact that the model is trained only on attack-free data, and detection depends on whether attack sequences produce reconstruction errors that differ sufficiently from normal behavior. This result shows that temporal information is useful for anomaly detection even when attack labels are not used during training. Therefore, both approaches can support SA by transforming low-level physical and traffic-level indicators into anomaly information that can be used as input for subsequent decision-making processes.

However, the study also presents some limitations. First, the available feature set is limited, which reduces the separability between normal and attack windows, especially for unsupervised detection. Secondly, the analysis is restricted to a single dataset and technology, so further validation is required to generalize the results to other environments. Finally, although the proposed models avoid using explicit device identifiers, the dataset is still associated with a fixed topology and particular devices.

Future work should explore the incorporation of additional discriminative features, the evaluation of alternative learning models, and the impact of different feature aggregation strategies and threshold selection methods on detection performance. In addition, explainability techniques could be incorporated to assess the contribution of physical-layer and traffic-level features to anomaly detection. Moreover, the models could be extended from binary detection to multi-class classification to identify specific attack types. Finally, the approach should be validated on additional datasets, technologies, and attack scenarios to assess its applicability to broader cyber-electromagnetic anomaly detection contexts.

\begin{credits}
\subsubsection{\ackname} 
This work has been co-funded by the European Union (EDF program; project ECYSAP EYE). Views and opinions expressed are however those of the author(s) only and do not necessarily reflect those of the European Union or the European Defence Fund. Neither the European Union nor the granting authority can be held responsible for them. 

\subsubsection{\discintname}
The authors have no competing interests to declare that are relevant to the content of this article.
\end{credits}
%
%
\bibliographystyle{splncs04}
\bibliography{references}

\end{document}